\documentclass[12pt]{JHEP3}

\makeatletter

\newbox\JHEP@outputbox
\gdef \@makecol {%
   \ifvoid\footins
     \setbox\@outputbox \box\@cclv
   \else
     \setbox\@outputbox \vbox {%
       \boxmaxdepth \@maxdepth
       \@tempdima\dp\@cclv
        \unvbox \@outputbox
       \unvbox \@cclv
       \vskip-\@tempdima
     }%
     \setbox\JHEP@outputbox \vbox {%
       \vskip \skip\footins
       \color@begingroup
         \normalcolor
         \footnoterule
         \unvbox \footins
       \color@endgroup
       }%
   \fi
   \xdef\@freelist{\@freelist\@midlist}%
   \global \let \@midlist \@empty
   \@combinefloats
   \ifvbox\@kludgeins
     \@makespecialcolbox
   \else
     \setbox\@outputbox \vbox to\@colht {%
       \@texttop
       \dimen@ \dp\@outputbox
       \unvbox \@outputbox
       \vskip -\dimen@
       \unvbox\JHEP@outputbox
       \@textbottom
       }%
   \fi
   \global \maxdepth \@maxdepth
}
\makeatother

\usepackage{epsfig}
\usepackage{array}

\def\AME{{\small\tt AMEGIC++}\ }

\title{SHERPA 1.{\boldmath$\alpha$}, a proof-of-concept version}
\preprint{CERN-TH/2003-284}
\author{Tanju Gleisberg\\ 
        Institut f\"ur Theoretische Physik, TU Dresden,
        01062 Dresden, Germany\\
        E-mail: \email{tanju@theory.phy.tu-dresden.de}}
\author{Stefan H{\"o}che\\ 
        Institut f\"ur Theoretische Physik, TU Dresden, 
        01062 Dresden, Germany\\
        E-mail: \email{hoeche@theory.phy.tu-dresden.de}}
\author{Frank Krauss\\ 
        Theory Division, CERN, CH-1211 Geneva 23, Switzerland\\
        {\rm and}\\
        Institut f\"ur Theoretische Physik, TU Dresden, 
        01062 Dresden, Germany\\
        E-mail: \email{krauss@theory.phy.tu-dresden.de}}
\author{Andreas Sch{\"a}licke\\ 
        Institut f\"ur Theoretische Physik, TU Dresden, 
        01062 Dresden, Germany\\
        E-mail: \email{dreas@theory.phy.tu-dresden.de}}
\author{Steffen Schumann\\ 
        Institut f\"ur Theoretische Physik, TU Dresden, 
        01062 Dresden, Germany\\
        E-mail: \email{steffen@theory.phy.tu-dresden.de}}
\author{Jan-Christopher Winter\\ 
        Institut f\"ur Theoretische Physik, TU Dresden, 
        01062 Dresden, Germany\\
        E-mail: \email{winter@theory.phy.tu-dresden.de}}

\abstract{
The new multipurpose event-generation framework {\tt SHERPA}, acronym
for {\tt S}imulation for {\tt H}igh-{\tt E}nergy {\tt R}eactions of
{\tt PA}rticles, is presented. It is entirely written in the
object-oriented programming language {\tt C++}. In its current form,
it is able to completely simulate electron--positron and unresolved
photon--photon collisions at high energies. Also, fully hadronic
collisions, such as, e.g., proton--anti-proton, proton--proton, or
resolved photon--photon reactions, can be described on the signal
level.}


\begin{document}
\section{Introduction}
\noindent
To a large amount, modern particle physics centres around accelerator
experiments, where high-energetic particles are brought to collision.
With rising energies, these interactions become more and more violent,
leading to an increasing number of particles being produced. To confront
the resulting experimental data with theoretical models, a systematic 
understanding of such multi-particle production processes is of paramount 
importance. A full, quantum-mechanically correct, treatment is, 
at the moment, out of reach. There are two reasons for this:

\noindent
First of all, there only is a limited understanding of the non-perturbative 
phase of QCD, or, in other words, of how colourless hadrons are built from 
the coloured quarks and gluons. This is especially true for phenomena 
such as hadronisation or for questions related to the impact of the partonic 
substructure of the colliding hadrons on the pattern of multiple interactions. 
In all such cases, phenomenological models for the transition from hadrons
to partons or vice versa have to be applied with parameters to be fitted. 
This clearly puts a constraint on a conceptual understanding of high-energy 
particle production processes. On the other hand, even considering the, in 
principle, well-understood perturbative phase of scattering processes alone, 
there are limits on our technical abilities to calculate all amplitudes 
that contribute to a given process. This is due to the fact that even at 
the tree-level the number of Feynman diagrams grows factorially with the 
number of particles involved. Moreover, at higher orders of the perturbative 
evolution new difficulties arise, which are connected for instance with the 
evaluation of multi-leg loop integrals.

\noindent
This failure necessitates other, approximate solutions, such as simulation 
programs. These event generators decompose the full scattering process into 
a sequence of different stages, which are usually characterised by different 
energy scales. The past and current success of event generators, like 
{\tt Pythia} \cite{Pythia} or {\tt Herwig} \cite{Herwig}, in describing a 
full wealth of various data justifies this decomposition intrinsic to all 
such programs. As a by-product, the decomposition of events into
distinguishable, more or less independent phases opens a path to test
the underlying assumptions on the dynamics of particle interactions at
the corresponding scales. These assumptions, in turn, can be modified
and new models can be included on all scales. This property turns
event generators into the perfect tool to bridge the gap between
experimental data and theoretical predictions. It renders them
indispensable for the analyses and planning of current and future
experiments.

\noindent
To meet the new challenges posed by the new experiments, for instance
Tevatron at Fermilab and especially LHC at CERN, the traditional event
generators {\tt Pythia} and {\tt Herwig}, so far programmed in {\tt
Fortran}, are currently being re-written in the modern,
object-oriented programming language {\tt C++}. Their new versions
will be called {\tt Pythia7} \cite{Pythia7} and {\tt Herwig++}
\cite{Herwig++}, respectively. The decision to re-write them from
scratch is based on two reasons:

\noindent
First, new features and models concerning the simulation of particle physics 
at the shifting energy frontier need to be included. In fact this still is 
an on-going issue also for the {\tt Fortran} versions (see for instance 
\cite{Moretti:2002eu,Sjostrand:2002ip}). 

\noindent
Furthermore, and maybe more importantly, there is a wide-spread belief that 
the old {\tt Fortran} codes cannot easily be maintained or extended. On top 
of that, the software paradigm of the new experiments has already shifted 
to object-orientation, more specifically, to {\tt C++} as programming 
language. On the other hand, by the virtue of being decomposed into nearly 
independent phases, the simulation of high-energy particle reactions lends 
itself to modularisation and, thus, to an object-oriented programming style. 
In this respect it is also natural to further disentangle management and 
physics issues in event generation. In fact, both {\tt Pythia7} and
{\tt Herwig++} will fully rely on the same management structure,
called {\tt ThePEG} \cite{ThePEG}.
It includes items such as the event record, mathematical functions, management 
functionalities, etc.. Using this common event-\-generation framework, 
{\tt Pythia7} and {\tt Herwig++} will just provide their respective, different 
modules for physics simulation, for instance the implementations of their 
hadronisation models. 

\noindent
In addition to these two re-writes of their older, {\tt Fortran}-based
counterparts, in the past few years a new event generator, called {\tt
SHERPA}, acronym for {\tt S}imulation for {\tt H}igh-{\tt E}nergy {\tt
R}eactions of {\tt PA}rticles, has been developed independently. From
the beginning, it entirely has been written in {\tt C++}, mainly due
to the same reasons already named above.
A number of paradigms have been the guiding principles in the
construction of this code:
\begin{enumerate}
\item Modularity:\\
      {\tt SHERPA} only provides the framework for event generation.
      The physics issues related to the various phases of event generation
      are handled by specific, physics-\-oriented modules. These modules,
      however, rely on a number of service modules that incorporate
      basic organisational, mathematical or physics tools, or information
      concerning the physics environment.
\item Separation of interface and implementation:\\
      Within {\tt SHERPA}, the specific physics modules are interfaced through
      corresponding (handler) classes, which are sufficiently abstract to
      support an easy inclusion of other modules with similar tasks.
\item Bottom-to-top approach:\\
      Before the interfaces (abstract handlers) are implemented, the
      corresponding physics module has been programmed and tested. This is
      especially true for modules like {\tt AMEGIC++} \cite{Krauss:2001iv},
      providing a full-fledged matrix-\-element generator for the evaluation 
      of multi-particle production cross sections, or {\tt APACIC++} 
      \cite{Kuhn:2000dk}, hosting a parton shower module. In general, 
      these modules can be used as stand-alone codes. They also can be
      implemented into other event-\-generation frameworks with minor
      modifications only, as long as some of the underlying mathematical and 
      physics tools are supplemented as well.
\end{enumerate}

\noindent
The goal of this publication is to give a brief status report of {\tt
SHERPA}'s first $\alpha$-version. It already incorporates enough
functionality to make {\tt SHERPA} a useful tool for a number of
physics applications.

\noindent
The outline of this paper is as follows: in Sec.~\ref{Overall} the overall 
generation framework is briefly introduced. This basically amounts to a discussion 
of how the framework and its physics modules are initialised, and how these 
modules are handed over to the actual event generation. Then, in the next 
two sections, Secs.~\ref{Tools} and \ref{Setup}, general tools for event 
generation, including for instance the event record, are presented as well 
as those modules that specify the physics environment (such as the physics model, 
beam spectra, or parton distribution functions), in which the simulation is 
performed. In the following, the implementation of some of the event phases 
reflecting different physics features will be briefly highlighted. The
discussion is commenced with describing the inclusion of hard matrix
elements for jet production etc. (Sec.~\ref{ME}) and for
heavy-\-particle decays such as, e.g., top-\-quark decays,
(Sec.~\ref{DEC}) into {\tt SHERPA}. Matrix elements are also 
needed for the simulation of multiple hard parton interactions in
hadronic collisions.
Hence, in Sec.~\ref{MI} a brief outlook will be given on how {\tt SHERPA} 
will describe such phenomena. In all cases mentioned above, the matrix
elements may give rise to configurations of jets to be fragmented by
the subsequent parton shower. A cornerstone of {\tt SHERPA} is the
implementation of an algorithm, which merges matrix elements and
parton showers respecting the next-to leading logarithmic accuracy of
the parton shower (for details on this algorithm, see \cite{CKKW}).
In Sec.~\ref{PS}, questions related to the inclusion of this algorithm
and the interplay with the parton shower inside the {\tt SHERPA}
framework are discussed. The quick tour through the event phases will
be finished in Sec.~\ref{Soft} with a discussion of issues related to soft 
QCD, e.g.\ hadronisation, beam jets, etc.. Finally, in Sec.~\ref{Finish}, 
conclusions will be drawn and a further outlook will be given.

\section{Overall event-generation framework\label{Overall}} 
\noindent
In {\tt SHERPA}, the various tasks related to event generation are
encapsulated in a number of specific modules. From a structural point
of view, the set-up of the event-generation framework condenses into
the problem to define the rules for the interplay of these modules and
to implement them. The flexibility to do so is increased by a
separation of the interfaces defining this interplay from the specific
modules -- the implementations of physics tasks\footnote{
Of course, this abstraction is to some extent limited by a kind of linguistic 
problem: in the implementation of the physics tasks, a choice has to be made 
on the terms in which the tasks are formulated. As a simple example consider 
four-momenta, clearly a basic ingredient of event generators. In {\tt ThePEG}, 
the choice has been made to represent them as five-vectors, where the fifth 
component denotes the mass related to the four-momentum; in contrast, 
in {\tt SHERPA} the representation is in terms of plain four-vectors. 
To use {\tt ThePEG} modules within {\tt SHERPA} requires a translation,
which in {\tt SHERPA} would be performed through the interface classes. The
objects defining the terms in which physics tasks are implemented
inside {\tt SHERPA} are accumulated in a namespace {\tt ATOOLS}, cf.\
Sec.~\ref{Tools}. Clearly, all other modules rely on these
definitions.}.
How this is realized within {\tt SHERPA} can be exemplified by the
hard matrix elements:

\noindent
There are two implementations, which can be used to generate hard partonic 
subprocesses. One of them is restricted to a list of analytically known 
$2\to 2$ processes, the other one is the multipurpose parton-level 
generator {\tt AMEGIC++}. However different they are, in the framework of 
event generation they have to calculate total cross sections for the hard 
subprocesses and they must provide single weighted or unweighted events. In 
{\tt SHERPA}, these functionalities of both modules are accessible through an 
interface, the {\tt Matrix\_\-Element\_\-Handler}. It naturally lives up 
to the intrinsic differences in these physics implementations. Without 
knowing any details about the realization of hard matrix elements in the 
modules, they can be plugged anywhere into the event-generation framework 
by means of this abstract handler class. To add another module concerned with 
hard partonic subprocesses, on the level of {\tt SHERPA} one would just
have to extend the corresponding methods of the {\tt
Matrix\_\-Element\_\-Handler} accordingly. This reflects a typical
object-oriented design principle.

\noindent
In general, such abstract handler classes encapsulate the specific physics 
implementations and are used to interface them with each other.
Further examples that have been implemented so far include the 
{\tt Beam\_Spectra\_Handler}, the {\tt ISR\_Handler}, the 
{\tt Hard\_Decay\_Handler}, the {\tt Shower\_Handler}, the 
{\tt Beam\_Remnant\_Handler} and the {\tt Fragmentation\_Handler}. They will
be described in the forthcoming sections.\\ 

\noindent
In many cases the underlying physics modules will require some 
initialisation before they can be used during event generation. 
Again, this can be exemplified by the hard matrix elements. In this case the 
initialisation basically consists of tasks like the set-up of matrix elements 
and phase-space integrators, and of the evaluation of total cross sections. 
They define the relative contributions of individual sub-processes in the 
overall composition of the hard process part inside the events. It is clear 
that such tasks have to be performed in an initialisation phase of an
event-generation run. During this phase, {\tt SHERPA} initialises the various 
physics modules selected by the user through the abstract handlers responsible 
for them. The specific set-up of a selected module will depend on
external, run-specific parameters, which are read-in from 
corresponding data files and managed by the same handler class. The 
initialisation sequence of these handlers and their physics modules is
organised by a {\tt SHERPA}-internal {\tt Initialization\_Handler},
which also owns the pointers to the handlers. To add new handlers for
completely new physics features, therefore, necessitates to modify and
extend this {\tt Initialization\_Handler}.\\

\begin{figure}[h]
\begin{tabular}{cc}
\includegraphics[width=7cm]{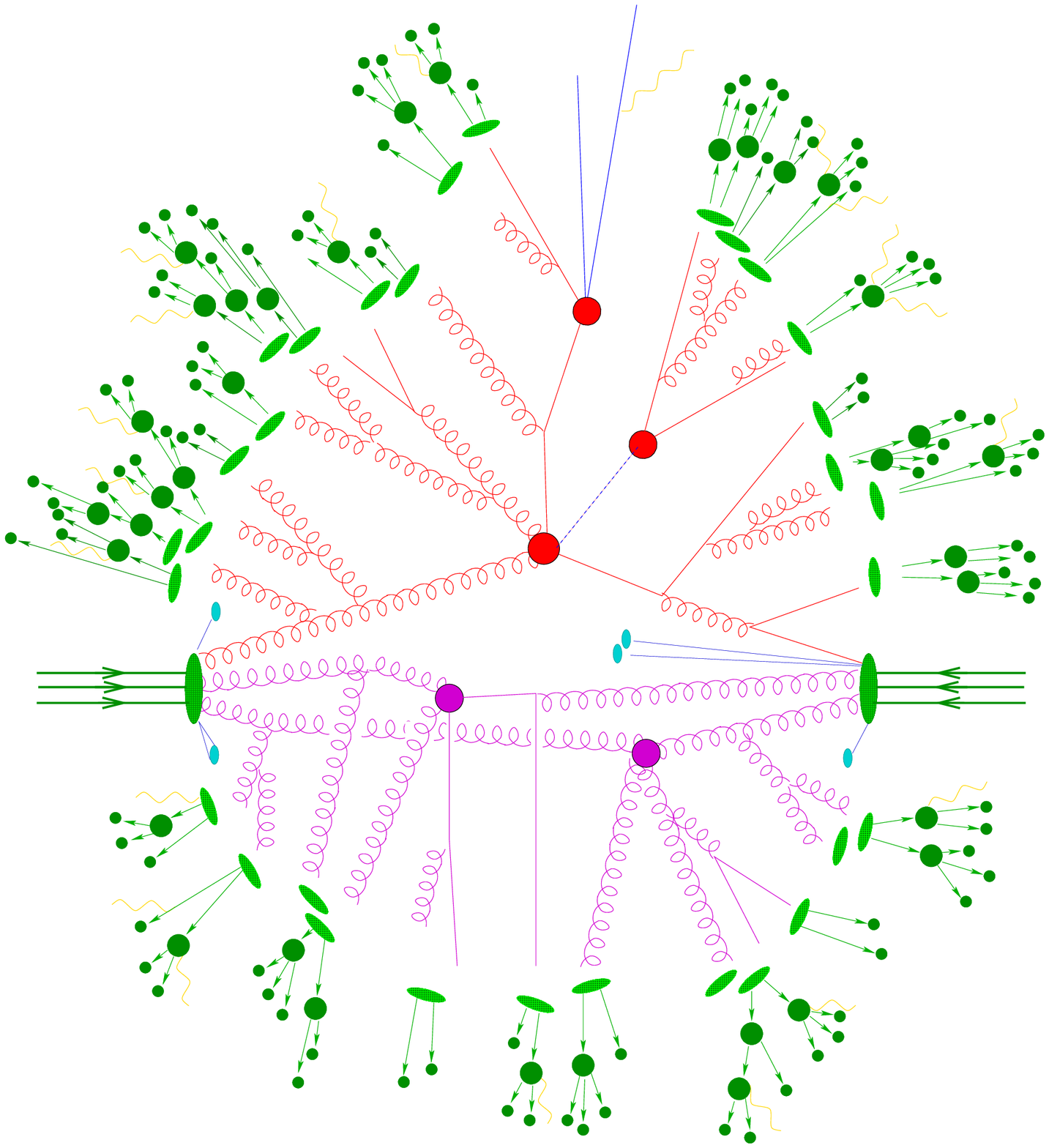}&
\includegraphics[width=8cm]{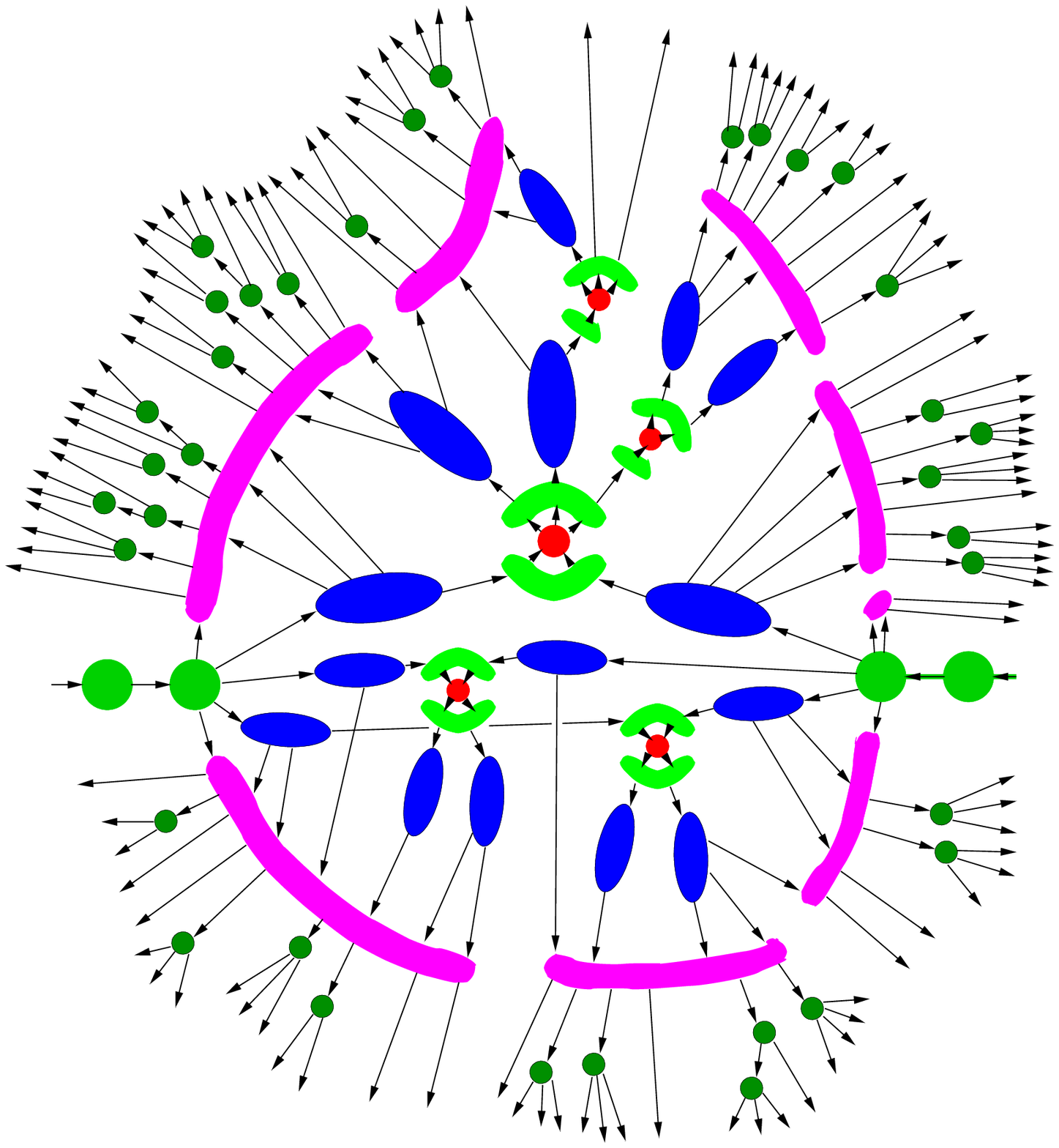}
\end{tabular}
\caption{\label{blobs}Pictorial representation of the event record. In the left
picture, a hadron--hadron collision is exhibited. Clearly, apart from the hard 
signal subprocess followed by hard decays of two heavy unstable
particles, it also contains two more hard parton interactions, all of
them shown as thick blobs.
The partons are dressed with secondary radiation as well, before the parton
ensemble is transformed into primary hadrons which then decay further.
On the right this is translated into the language of {\tt Blob}s.
Here, each hard matrix-element {\tt Blob} (red) is equipped with
merging {\tt Blob}s (green) in the initial and final state which
define initial conditions for the parton shower. All extra partons
emitted during the shower evolution are combined in individual
shower {\tt Blob}s (blue). In the hadronisation {\tt Blob}s (magenta)
colour singlet chains formed by incoming partons are translated into
primary hadrons which might decay further. Each such hadron decay is
represented by an extra {\tt Blob}.}
\end{figure}

\noindent
Having initialised the interfaces to the physics modules, the {\tt SHERPA}
framework is ready for event generation. As already stated before, the 
individual events are decomposed into separate phases. This decomposition 
is reflected by {\tt SHERPA}'s program structure in the following way:
an {\tt Event\_Handler} object manages the generation of one single event 
by having a list of various {\tt Event\_Phase\_Handler}s acting on the 
expanding event record. This process of event generation is formulated 
in terms of particles connecting generalised vertices, coined blobs. These 
{\tt Blob}s in turn reflect the space-time structure of the event, each 
of them has a list of incoming and outgoing particles. In other words, the 
blobs are the nodes, the particles are the connecting lines of a network.
For a pictorial example, confronting a simple hadron--hadron event
with its representation through {\tt Blob}s, cf.\ Fig.~\ref{blobs}.
An event thus can be represented as a list of {\tt Blob}s, which in turn
forms {\tt SHERPA}'s event record. The {\tt Event\_Phase\_Handler}s
act on this list, by either modifying the {\tt Blob}s themselves or by
adding new {\tt Blob}s or by subtracting unwanted ones. For event
generation, the list of {\tt Event\_Phase\_Handler}s is tried on the
list of {\tt Blob}s until no more action is possible, i.e.\ until none
of the individual {\tt Event\_Phase\_Handler}s finds an active {\tt
Blob} it can deal with.
To illustrate this, consider the following simple example:
\begin{itemize}
\item First of all, a yet unspecified blob of the type
      {\it``Signal Process''} is added to the so far empty {\tt Blob} 
      list. Iterating with the list of {\tt Event\_\-Phase\_\-Handler}s the 
      {\tt Signal\_\-Processes} phase deals with the single
      unspecified active {\tt Blob}, inserting a number of incoming
      and outgoing partons through the {\tt Matrix\_\-Element\_\-Handler}.
\item In the next iteration of the {\tt Event\_\-Phase\_\-Handler}s,
      the {\tt Jet\_Evolution} phase steps over this {\tt Blob} and
      adds parton showers to it. To this end, some {\it``ME PS Interface''}
      {\tt Blob}s are added as well as some {\tt Blob}s for the initial-
      and final-state parton shower, signified by the types 
      {\it``IS Shower''} and {\it``FS Shower''}, respectively.
      Assuming that an $e^+e^-$ annihilation into hadrons is simulated, 
      the {\it``IS Shower''} {\tt Blob}s have one incoming and one
      outgoing electron each, and, maybe, some outgoing photons as well.
      The {\it``Signal Process''} as well as the {\it``ME PS Interface''}
      {\tt Blob}s are switched to passive by this phase.
\item The {\tt Hadronisation} phase selects out the shower {\tt Blob}s
      for the transition of partons into hadrons. First the 
      {\tt Beam\_\-Remnant\_\-Handler} has to fill {\it``Beam Remnant''}
      and {\it``Bunch''} {\tt Blob}s. In the toy example, both,
      however, have a simple structure with one incoming and one 
      outgoing electron each. Now, the {\tt Fragmentation\_Handler} comes into 
      play, adding more blobs of the type {\it``Fragmentation''} with a 
      number of incoming partons and a number of outgoing primary hadrons. 
      All {\tt Blob}s apart from the {\it``Fragmentation''} ones would
      be switched to passive now, leaving the outgoing primary hadrons 
      to be decayed. These decays would be represented by more 
      {\tt Blob}s of the type {\it``Hadron Decay''}.
\end{itemize}
The structure elucidated above allows for nearly arbitrary mixtures in the 
composition of an event. For example, through the action of the {\tt
Jet\_Evolution} phase the parton shower could in principle alternate
with a sequence of hard decays on the parton level, or it could even
be invoked in the decay of a heavy hadron.

\begin{figure}[h]
\includegraphics[width=16cm]{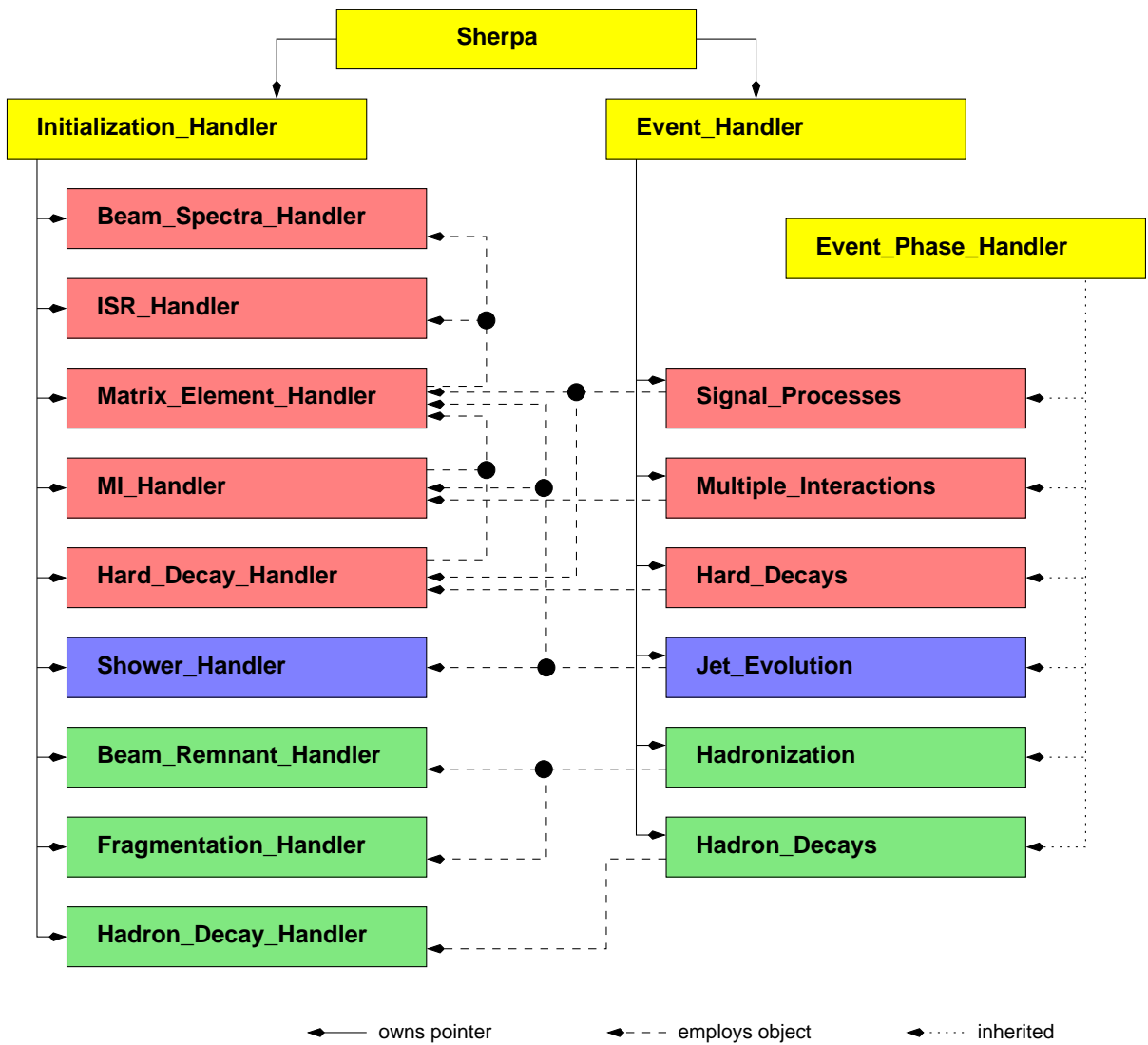}
\caption{\label{phases}The {\tt Event\_Phase\_Handler}s and their
interfaces, all of which are implemented up to now in {\tt SHERPA}.}
\end{figure}

\noindent
In Fig.~\ref{phases} the {\tt Event\_Phase\_Handler}s implemented so
far and their connections to various interfaces are exhibited. 
\section{Tools for event generation\label{Tools}}
\noindent
In {\tt SHERPA}, the basic infrastructure for event generation, which is used 
by other modules, is centralised in a separate package, called {\tt ATOOLS}. 
It contains management, mathematics, and physics tools. 

\noindent
The organisational tools include, among others, classes to read-in input data, 
and to provide parameters and objects that must be globally accessible. During 
the initialisation of the {\tt SHERPA} environment this data-container
class is instantiated as a global object, which is filled and accessed
by the other modules in due course. Therefore, if a potential user
wants to include more objects that are needed in very separate corners
of the total framework, he or she would have to include these objects
into this class {\tt Run\_Parameters}. Of course, the corresponding
access methods have to be provided there as well.
{\tt SHERPA} offers the possibility to specify a large amount of parameters 
for a run without recompiling. To enhance the transparency of the read-in 
procedure and to contribute to its intuitive understanding, the
variables might be contained in different, user-specified data files
in the following fashion:
\begin{verbatim}
  KEYWORD = Value .
\end{verbatim}
Within the code, default values can be given for the parameters
connected to the keywords. An example defining, e.g.\ the physics
model, and declaring the Standard Model as the default choice, reads:
\begin{verbatim}
  Data_Read _dataread(path,file);
  std::string model = _dataread.GetValue("MODEL",std::string("SM"));
\end{verbatim}
In its instantiation, the {\tt \_dataread}-object is given the path
and the file name for the read-in procedure.\\

\noindent
A second group provides mathematical service classes, including:
\begin{itemize}
\item a representation of three- and four-vectors; 
\item a class for real or complex matrices;
\item a representation of Lorentz-transformations (boosts and rotations); 
\item abstract definitions of functions or grids which can be integrated 
      or inverted;
\item a class for simple histograms and operations on them;
\item the random number generator.
\end{itemize}
This group of objects defines the mathematical terms in which {\tt SHERPA} 
generates events.\\

\noindent
The basic physics terms are also part of the {\tt ATOOLS} package and cover a 
wide range of applications. In the following, some of the corresponding basic 
classes will be briefly described:
\begin{itemize}
\item Particles are described by some, in principle, unchangeable
      characteristics:
      their quantum numbers, their mass and width, etc.. All these
      properties are contained in a {\tt Flavour} object. Within {\tt
      SHERPA}, also pseudo-flavours, for instance ``{\tt jet}'', are
      available. Hence, a {\tt Flavour} object might serve as a container
      for other {\tt Flavour}s. In {\tt SHERPA} the particles
      and their properties are collected in two data files, {\tt Particle.dat}
      and {\tt Hadron.dat}. A typical line in these files looks like:
      \\\\
      \hspace*{-1.4cm}
      \begin{minipage}[h]{19cm}
      \begin{verbatim}
      kf  Mass  Width  3*e  Y  SU(3)  2*Spin  maj  on  stbl  m_on  Name
      1   .01   .0     -1   -1   1      1      0   1    1     0    d
      \end{verbatim}
      \end{minipage}
      \noindent
      Apart from the mass, width and spin, the electrical charge, the
      third component of the weak iso-spin, and the ability to
      participate in strong interactions are defined. In addition, for
      fermions, the user should provide information whether a specific
      {\tt Flavour} describes Majorana particles or not. Also, information 
      has to be provided, whether individual particles should be
      included at all, whether they are stable or not, and whether
      their mass should be taken into account in matrix-element
      calculations\footnote{It should be mentioned here that this 
      mass enters in the phase space and in the propagators. For the
      Yukawa couplings these masses, if switched on, serve as default
      value, but can be overwritten during the initialisation of the
      physics models.}. Finally, the particles' names should be
      defined as well in a form that will show up in the event record.
\item In some cases, the user might wish to have, e.g., the
      matrix-element generator(s) to calculate the width of a {\tt
      Flavour}, thus overwriting the one given in
      {\tt Particle.dat}. To this end, another data file, by default
      called {\tt Decays.dat}, might be read-in. Then, for the
      corresponding particles, decay tables are constructed 
      and evaluated. They are implemented as {\tt Decay\_Table} objects.
\item The particles, which finally show up in the generated event, are
      represented through a class {\tt Particle}. In addition to the
      data objects specifying its properties, the {\tt Particle}s are
      characterised by their four-momenta, by the vertices ({\tt Blob}s)
      in which they are created or end, and by the flow of quantum
      numbers associated with them, such as colour.
\end{itemize}
In addition to the classes outlined above, the {\tt ATOOLS} package includes classes which 
define some physics observables or which can be used to select events. These {\tt Selector} classes
are also needed for the integration over the phase space of the final state in hard subprocesses. 
One of them is providing a definition of jets according to the $k_\perp$- (or Durham-) algorithm
\cite{Catani:1991hj} in various collision types. It is of special importance for the 
{\tt SHERPA} package, since it is used for the merging procedure of matrix elements and the 
parton shower.

\section{Physics set-up\label{Setup}}
\noindent
In this section those packages are presented that define the overall physics set-up.
Clearly, this contains the specification of the physics model, in which cross sections 
are calculated or events are generated. Such a physics model defines the set of particles 
in it as well as most of their properties, including their mutual interactions. Equally 
important is a declaration of which type of process is discussed. Basically this amounts
to a definition of incoming beams and their structures, both in terms of their respective
energy spread and in terms of their eventual partonic substructure, which can be parametrised 
by parton distribution functions. In the following, therefore, the packages
{\tt MODEL}, {\tt BEAM}, and {\tt PDF} are briefly introduced. Within {\tt SHERPA} they
define the physics model, the structure of the incoming beams and the eventual inner 
structure of the colliding particles, respectively.\\

\noindent
The package {\tt MODEL} encapsulates abstract structures to specify
arbitrary parameter sets of physical models, e.g.\ coupling constants, 
Yukawa masses, decay widths, etc.. For a certain physical model,  
for instance the Standard Model or its minimal super-symmetric extension,
all parameters are represented by a {\tt Model} object derived from 
the abstract base class {\tt Model\_Base}. This base class and its explicit
instances mainly serve as containers and handle the input and the access 
to the parameters. The main ingredients of this class are lists of
four standard parameter types: 
\begin{itemize}
\item {\tt ScalarNumber} for integer constants, 
\item {\tt ScalarConstant} for floating point (double precision) constants, 
\item {\tt ScalarFunction} for real single-parameter functions, derived from the abstract class 
      {\tt ATOOLS::Function\_Base}, and
\item {\tt ComplexMatrix} for a matrix of complex floating point 
      (double precision) constants. 
\end{itemize}
Examples of parameters, which could be contained in the lists, are the 
number of extra dimensions, $\alpha$ in the Thomson limit, the running 
strong coupling constant $\alpha_s$, and the CKM-matrix, respectively.
Each parameter is mapped on a name string, which is used for all references 
on the parameter. A code example for the insertion of such a pair of name
and parameter into the list of scalar constants reads
\begin{center}
\begin{verbatim}
   p_constants->insert(std::make_pair(std::string("ALPHAQED(0)"),
                                      1./137.03599976));
\end{verbatim}
\end{center}
To access parameters, the class {\tt Model\_Base} defines a function
for each parameter type, for instance the constant {\tt "ALPHAQED(0)"} 
can be re-obtained through a call of
\begin{center}
\begin{verbatim}
   ScalarConstant("ALPHAQED(0)");
\end{verbatim}
\end{center}

\noindent
There are two typical situations for setting the parameters of a certain
model. First, they can be simply read-in from a file, which by default 
is called {\tt Model.dat}. As a second possibility, {\tt Model\_\-Base} 
is equipped with a pointer to a {\tt Spectrum\_\-Generator\_\-Base} object. 
This object provides an abstract interface to external spectrum generators
with methods to read-in input parameters, to deduce the particle spectrum
and to calculate the other parameters of this model. So far, interfaces to 
the {\tt Fortran} codes {\tt Hdecay} \cite{Djouadi:1998yw} and {\tt Isajet} 
\cite{Baer:1999sp} have been constructed. They are instances of the abstract 
base class {\tt Spectrum\_\-Generator\_\-Base} and they are called
{\tt Hdecay\_\-Fortran\_\-Interface} and {\tt Isajet\_\-Fortran\_\-Interface},
respectively. To include more of these generators, a user would have
to derive such an interface class and provide methods to read-in
the input parameter set, to calculate the other parameters and to
modify the particle spectrum accordingly. It should be noted that for the 
inclusion of new particles, also the class {\tt Flavour} would have to
be extended correspondingly\footnote{Using the new accord on a generic 
interface structure for spectrum generators, \cite{Skands:2003cj}, 
the task to inherit new instances of the {\tt Spectrum\_\-Generator\_\-Base}
will be substantially alleviated.}.\\

\noindent
Within {\tt SHERPA} the original beams of a specific collider are
treated in two different stages in order to extract the partonic initial 
states for the hard interactions. In the first step, the incoming beams
at a certain energy, the nominal energy of the collider, are
transfered into bunches of interacting particles, which have an energy
distribution, and whose momenta are distributed collinearly w.r.t.\ 
the original beams. Two options are currently implemented:
the beams can either be monochromatic, and therefore need no extra
treatment, or, for the case of an electron collider, Laser backscattering 
off the electrons is supported. This mode leads to photon bunches with a 
certain energy and polarisation distribution. In a second step, possible 
substructures of the bunch particles are taken into account, as well as 
ordinary initial state radiation. This task is achieved by means of parton 
distribution functions (PDFs) or simple structure functions for the case of 
electron ISR. 

\noindent
As an illustrative example, consider the case of resolved photon
interactions at an electron collider. As stated above, by Laser backscattering 
the incoming electrons can be ``transformed'' into photons
distributed in energy and polarisation depending on the parameters
chosen for the incoming electron beam and the Laser. This corresponds
to the first step. In the second step, these photons have a
partonic substructure described by an appropriate photon PDF
defining the probability to find a certain parton flavour at the scale 
$Q^2$ and the energy fraction $x$ inside the photon. \\

\noindent
The first stage is hosted in the module {\tt BEAM}, housing all classes 
that are employed to generate beam spectra. The handler class to access 
different beam-manipulation strategies is {\tt Beam\_Spectra\_Handler}. Before coming 
into full effect during integration or event generation, this handler 
initialises a suitable treatment ({\tt Beam\_Base}s) for both beams and 
uses them to generate corresponding weights, i.e.\ energy distributions. At 
the moment, all outgoing bunch particles are still collinear to the
incoming beams, but this is going to change in the future, by adding 
transversal boosts to the kinematics. Up to now two types of 
{\tt Beam\_Base}s are supported: {\tt Monochromatic} beams, and the 
generation of photon beams via {\tt Laser\_Backscattering}.
For the latter one the parametrisation of \cite{Zarnecki:2002qr} is supplied in addition 
to a simple theoretical ansatz. To flatten 
out the peaks in the energy distribution of the produced photons, additional
phase-\-space mappings have been introduced, which are located in the module 
{\tt PHASIC++} and come to action as further channels in a multi-\-channel
phase-\-space sampling \cite{Berends:1994pv} also implemented there. For 
more details, cf.\ Sec.~\ref{ME}. To implement any new beam treatment, 
such as, e.g., Beamstrahlung, a corresponding instance of the class 
{\tt Beam\_Base} has to be provided. In addition, the construction of extra 
phase-\-space mappings might become mandatory.  
 
\noindent
The second stage, i.e.\ the handling of initial state radiation or partonic 
substructures, is located in the {\tt PDF} module. The handler class steering the 
selection of PDFs or structure functions of bunch particles is the {\tt PDF\_Handler},
instantiating a suitable {\tt PDF\_Base} object and returning a pointer to it. 
So far, a structure function for electrons (that can handle charged leptons 
in general), a photon PDF and various proton structure functions are 
available. The list of proton PDFs covers: a {\tt C++} version of MRST99
\cite{Martin:1999ww}, the {\tt Fortran} CTEQ6 PDF \cite{Pumplin:2002vw}, 
and the set of {\tt LHAPDF}s \cite{LHAPDF}. 
The two {\tt Fortran} pieces are encapsulated by the two classes 
{\tt CTEQ6\_Fortran\_Interface} and {\tt LHAPDF\_Fortran\_Interface}. 
For the case of photon bunches, the only structure function implemented
is the GRV (LO) parton density \cite{Gluck:1991jc}, again framed by a 
C++ class, {\tt GRVph\_Fortran\_Interface}. 
Having selected and initialised all required PDFs the {\tt PDF\_Base} 
objects are handed over to the {\tt ISR\_Handler} via pointers
to two {\tt ISR\_Base} objects. If no ISR treatment is necessary for 
a beam the {\tt ISR\_Base} is instantiated as an {\tt Intact} object, else 
a {\tt Structure\_Function} object is instantiated, which possesses a pointer 
to the corresponding {\tt PDF\_Base}. At first glance this construction
looks quite over-engineered, however, it allows for a straightforward
implementation of possible multi-parton structure functions, which one 
would possibly like to use to correctly account for multiple interactions.  
To efficiently sample initial state radiation or parton distributions, and
similar to the beam treatment, qualified phase-space
mappings have been constructed, taking into account the peak structure
of the corresponding distributions. It is also worth noting that the PDFs are
handed over to the {\tt Shower\_Handler} in order to facilitate the
backward evolution of initial-\-state parton showers, see Sec.~\ref{PS}.
\section{Matrix elements and phase space integration\label{ME}}
In the {\tt SHERPA} framework, hard matrix elements occur in different
phases of event generation, i.e.\ in the generation of the (hardest) signal
process, in the decay of heavy unstable particles, or during the simulation
of multiple parton interactions. This is reflected by the appearance of 
different {\tt Event\_Phase\_Handler}s during event generation. In fact,
event generation starts with an empty list of blobs. The first blob to be 
filled by the {\tt Signal\_Processes} event phase is, obviously, for the 
partonic signal process. This event phase, like the other ones, such as
{\tt Hard\_Decays} and {\tt Multiple\_Interactions}, owns a pointer to an
appropriate handler for the matrix elements.

\noindent
As briefly mentioned before, {\tt SHERPA} currently incorporates
two modules concerned with matrix elements for hard partonic subprocesses.
These modules are interfaced through the {\tt Matrix\_Element\_Handler},
which in turn possesses public methods for the set-up of the calculation 
framework (physics model, beam spectra, PDFs, construction of suitable,
process- and framework-dependent integration channels), for the evaluation 
of total cross sections, and for the generation of single events. These 
tasks as well as some management issues (number and flavour of partons, etc.)
look very similar on an abstract level, and in fact, the corresponding methods
just call their counterparts in the specific matrix element realisation.
There is one difference, however, in these modules. The analytically known
$2\to 2$ processes incorporated in the module {\tt EXTRA\_XS} provide
the colour structure of individual parton configurations through specific 
methods. {\tt SetColours} defines this structure in terms of the
external four-momenta, whereas {\tt Colours} returns the colour structure. 
In {\tt AMEGIC++} things are not so easy. In fact, in {\tt SHERPA} the
colour structure of an $n$-parton configuration is reconstructed by
backward clustering, which is guided by the individual Feynman diagrams,
cf.\ Sec.~\ref{PS}. This algorithm allows, in principle, to reconstruct
colour flows for any multi-parton configuration in a leading-log large-$N_c$
scheme for any parton level generator. The only ingredient that has to
be delivered by the parton-level generators is a representation of Feynman 
diagrams in terms of binary trees. Therefore, {\tt AMEGIC++} 
provides methods to access the amplitudes. This difference is also reflected 
in the {\tt Matrix\_Element\_Handler}. It allows to either directly 
access the class responsible for the hard $2\to 2$ subprocesses in the case of
{\tt EXTRA\_XS} or to extract individual Feynman diagrams from
{\tt AMEGIC++}.\\

\noindent 
The library {\tt EXTRA\_XS} supplies a list of simple $2\to 2$ processes at 
leading order and their analytically known differential cross sections. Thus 
it allows for a fast evaluation of such processes. At present it includes 
all $2\to 2$ QCD and Drell-Yan processes with massless partons. Furthermore, 
it is employed for the determination of the initial colour configuration for 
the parton shower during event generation. When \AME is used as 
signal generator, this applies after an appropriate backward clustering, 
cf.\ Sec.~\ref{PS}. 

\noindent 
Within {\tt EXTRA\_XS} each process object is inherited from the base class 
{\tt XS\_Base}, which contains the basic ingredients for a $2\to2$ signal 
generator. This amounts to methods providing the particle types, the total and 
differential cross section of the process, and to methods that allow the generation 
of single parton-level events and the determination of their colour structure. In 
the set-up of such an {\tt XS\_Base} the overall physics model, the beam spectra and 
the ISR strategy have to be handed over as well. The latter information is
employed to select adequate initial state channels for the phase-space 
integration (see below). Since only $2\to2$ processes are taken into account 
within {\tt EXTRA\_XS}, its final state part boils down to simple hard wired 
\texttt{S}-, \texttt{T}- and \texttt{U}-channel integrators. According to 
its specific purpose, an {\tt XS\_Base} object may either correspond to a 
single $2\to 2$ process represented by an instance of the class 
{\tt Single\_XS} or to a set of processes represented by the container class 
{\tt XS\_Group}. However, if a user wants to set up his own process, he or 
she has to derive it from {\tt Single\_XS} and to define all its process-specific
properties, such as the colour structure of the particles involved, the 
differential cross section or the final state channels. The overall 
interface from {\tt EXTRA\_XS} to the {\tt SHERPA} framework is the special 
{\tt XS\_Group} called {\tt Simple\_XSecs}, which can be accessed through 
the {\tt Matrix\_Element\_Handler} and serves as a signal generator. 
This class also contains methods to read-in user-defined run-specific 
subprocesses and to select and initialise the corresponding 
{\tt XS\_Base}s.\\

\noindent
{\tt AMEGIC++} is {\tt SHERPA}'s prefered multipurpose matrix-element 
generator concerned with the production and evaluation of matrix elements 
for hard processes in particle collisions at the tree-level. A manual for 
the current version 2.0 is in preparation, superseding an older one,
\cite{Krauss:2001iv}. This new version now also covers the 
full Minimal Supersymmetric Standard Model (MSSM) 
\cite{Haber:1984rc,Rosiek:1989rs} and the ADD model 
\cite{Arkani-Hamed:1998rs} of large extra dimensions; for details 
concerning the implementation of the latter one, see \cite{Gleisberg:2003ue}.

\noindent
In its instantiation, {\tt AMEGIC++} is equipped with pointers to a 
{\tt Model\_Base} object, to a {\tt Beam\_Spectra\_Handler} and to an 
{\tt ISR\_Handler} object. The first one supplies all coupling constants 
and model specific parameters that allow {\tt AMEGIC++} to construct a list 
of all available Feynman rules, i.e.\ vertices, for the chosen physical model. 
They are represented through objects of the type {\tt Single\_Vertex}, which
possess pointers to a {\tt Lorentz\_Function} and a {\tt Color\_Function} 
object accounting for the intrinsic Lorentz and $SU(3)$ colour structure 
of the vertex. This is nicely exemplified by the triple gluon vertex:
\begin{verbatim}        
  Kabbala kcpl0 = -g3;
  Kabbala kcpl1 = kcpl0; 
  
  for (short int i=0;i<3;i++) 
    vertex[vanz].in[i] = Flavour(kf::gluon);
        
  vertex[vanz].cpl[0]        = kcpl0.Value();
  vertex[vanz].cpl[1]        = kcpl1.Value();
  vertex[vanz].cpl[2]        = 0.;
  vertex[vanz].cpl[3]        = 0.;
  vertex[vanz].Str           = (kcpl0*PR+kcpl1*PL).String();

  vertex[vanz].ncf           = 1;
  vertex[vanz].Color         = new Color_Function(cf::F);     
  vertex[vanz].Color->SetParticleArg(0,2,1);     
  vertex[vanz].Color->SetStringArg('0','2','1');     

  vertex[vanz].nlf           = 1;
  vertex[vanz].Lorentz       = new Lorentz_Function(lf::Gauge3);     
  vertex[vanz].Lorentz->SetParticleArg(0,1,2);     

  vertex[vanz].on            = 1;
  vanz++;
\end{verbatim}
\noindent
To extend {\tt AMEGIC++} and incorporate new interaction models, a
potential user would just have to derive a corresponding class from the
{\tt Interaction\_Model\_Base} class and to fill it with suitable vertices.

\noindent
Having specified a process or a group of processes to be evaluated, 
{\tt AMEGIC++} then constructs all Feynman diagrams by matching the set 
of vertices onto topologies generated beforehand. These amplitudes are 
translated into helicity amplitudes, which are subject of various 
manipulations, all aiming at a reduction of the calculational cost of the 
entire computation. As a further step {\tt AMEGIC++} analyses all individual 
Feynman diagrams and, according to their phase-\-space singularities, it 
automatically generates appropriate phase-\-space mappings for the integration 
over the final state. For more details on the multi-\-channel integration, see
below. The integration channels as well as the helicity amplitudes 
are stored as library files that have to be compiled once and are linked to 
the main program. The by far most convincing features of the {\tt AMEGIC++}
module are its robustness and flexibility. The package offers the evaluation 
of arbitrary processes\footnote{{\tt AMEGIC++} has proved to work for up 
to six final state particles \cite{Costas6Ferms}.} 
in the Standard Model, and in two of its extensions, the MSSM and the ADD model. \\

\noindent
The tools for phase-space integrations, i.e.\ simple integration channels,
building blocks for complex phase-space mappings and the full set of 
multi-channel integration \cite{Berends:1994pv} routines are hosted in the 
package {\tt PHASIC++}. It is used by {\tt AMEGIC++} as well as by the 
simple matrix elements located in the {\tt EXTRA\_XS} package. If 
needed, it can be adjusted in a straightforward fashion for usage by 
any other matrix element generator. The only thing, one would have to 
do, is to provide information about or to directly construct the 
channels for the final state part. Both strategies are realized by
{\tt EXTRA\_XS} and by {\tt AMEGIC++}, respectively. In the latter case,
the class responsible for the construction of the full final-state
multi-channel integrator is the {\tt Phase\_Space\_Generator}, individual
channels are constructed by the {\tt Channel\_Generator} through a mapping of 
the Feynman diagrams onto the {\tt Channel\_Elements} supplemented by
{\tt PHASIC++}.

\noindent
Apart from the matrix-\-element-\-specific final-\-state channels, during 
the phase-\-space integration one might have to sample over all initial-\-state 
configurations. Within {\tt SHERPA} initial states on the parton level are 
constructed from the incoming beams in two steps. First, the beam particles 
might be transformed into other particles (such as electrons into photons 
through Laser backscattering) or may experience some smearing (such as
electrons through Beamstrahlung). The resulting particles, which may or may 
not have an energy distribution, might have a resolved partonic 
substructure parametrised by PDFs or they might experience additional 
initial state radiation, which can also be parametrised by a PDF-like 
structure. To guarantee optimal integration performance, one has to analyse 
the emerging energy distributions in each of the two steps and flatten them 
out. This results in up to two more multi-\-channel mappings, one for the 
beam centre-\-of-\-mass system, and one for the parton-\-level centre-\-of-\-mass 
system. Both systems currently are defined through the boost relative to 
their ancestors and by their respective centre-\-of-\-mass energy squared. In 
the near future, also transversal boosts of the subsystems will be included. 
This, however, is a straightforward extension of existing code.
\section{Decays of unstable particles\label{DEC}}
\noindent
Decays of heavy unstable particles during the generation of an event are treated
by a specific {\tt Event\_Phase\_Handler} called {\tt Hard\_Decays}. This
handler owns, not surprisingly, an interface to matrix elements 
responsible for the description of such decays on the parton level. Again,
this interface, the {\tt Hard\_Decay\_Handler}, is separated from the physics
implementation, namely the matrix elements. Currently, only the matrix 
elements of {\tt AMEGIC++} are accessible through this interface. 

\noindent
At the moment, heavy unstable particles are produced by hard matrix 
elements only, i.e.\ through the actions of the following event phases:
{\tt Signal\_\-Processes}, {\tt Hard\_\-Processes} and 
{\tt Multiple\_\-Interactions}. While processing each of these phases,
it is checked whether unstable particles emerge. If this is the case, their 
respective decay channel and the effective mass of this decay are determined. 
The decay channel is selected by invoking the {\tt Hard\_\-Decay\_\-Handler},
which provides a mapping of particles to decay tables and the corresponding
matrix elements for each decay channel. Hence, a pointer to this interface
is a member of all the event phases above. The effective mass is distributed 
according to a Breit--Wigner function, the method for this resides in the 
{\tt Particle} object itself. Fixing the decay channel before the mass is
determined ensures that the correct, initialised branching ratios are 
recovered. In principle, this also allows for using tree-level decay 
kinematics as supplemented by, e.g., {\tt AMEGIC++} together with higher order 
branching ratios\footnote{Such a procedure might seem somewhat inconsistent. 
However, using loop-corrections for, say, two-body decays, basically amounts 
to a specific choice of scale of the coupling constant(s) involved. In this 
sense, inconsistencies are due to different choices of scale, which could be 
fixed and compensated for in the corresponding vertices.}. After all masses 
are fixed, the four-momenta of all particles emerging in the corresponding 
hard subprocess (all particles leaving the blob) are shifted to their new 
mass-shell accordingly. This induces some minimal modifications of the
energy-momentum relations of the particles and might affect the mutual
respective angles. However, the four-momentum of the total system stays
fixed. Eventually, after some jet evolution took place, the unstable 
particles are decayed, maybe giving rise to more unstable particles or new 
jets and, thus, triggering more actions of the {\tt Hard\_Decays} or 
{\tt Jet\_Evolution} phase.
 
\noindent
At the moment, the procedure outlined above is being implemented and tested. 
In its current, minimal form, two issues have not been tackled:
\begin{itemize}
\item In principle, attaching secondary radiation to hard decays leads to
      multi-scale parton showers \cite{Gieseke:2002sg}, which act in the 
      following way: In a first step the parton shower evolves the parton 
      configuration down to scales comparable to the width of the decaying 
      particles. Then, these particles decay, eventually starting an initial 
      and a final state parton shower, which have to be matched with the 
      preceeding one. Finally, the emerging parton ensemble is evolved down 
      to the next decay or the infrared scale. An implementation of this 
      procedure is straightforward in the {\tt SHERPA} framework.
\item Furthermore, spin correlations in the fashion of \cite{Richardson:2001df}
      should be added. The underlying idea is as follows. When
      decays of heavy unstable particles are treated in the way outlined
      above, implicitly some narrow width approximation has been used. In 
      fact, this inherent assumption only allows to cut the propagators of 
      the unstable particles\footnote{In other words, if the decaying 
      particles' width becomes large, all processes, i.e.\ also the 
      ``continuum'' or background, contributing to the same final state have 
      to be taken into account.}. Under the narrow width approximation, one 
      can decompose the propagator into a sum over physical polarisation 
      states. The polarisations of a number of outgoing particles produced 
      in one interaction, however, are correlated, and this correlation 
      propagates to a correlation in the kinematical distribution of the 
      decay products.
\end{itemize}
\section{Multiple interactions\label{MI}}
\noindent
Multiple interactions are handled within the {\tt SHERPA} framework by the 
{\tt Event\_\-Phase\_\-Handler} called {\tt Multiple\_Interactions}. Given a 
{\tt Blob} list, which already contains the signal process, it adds one by one
hard $2\to 2$ subprocesses, according to an ordering in the transverse momentum
$p_\perp$ of the outgoing particles. The initial conditions for this sequence of 
parton interactions are determined by the signal process. However, it might happen 
that the signal process contains more than two outgoing particles and, thus, 
the definition of $p_\perp$ is ambiguous. Then, the backward clustering 
already employed to create an interface from the signal process to the parton 
shower (see Sec.~\ref{PS}) defines the corresponding $2\to 2$ process. The
sequence of further partonic $2\to 2$ interactions results in new {\tt Blob}s,
each of which experiences its own shower evolution through the action of
the {\tt Jet\_Evolution} event phase. 

\noindent 
To create the additional hard subprocesses, the {\tt Multiple\_Interactions}
phase employs a {\tt MI\_Handler}, the interface to the new module 
{\tt AMISIC++}. This module is concerned with the generation of hard 
underlying events similar to how this is simulated in {\tt Pythia} 
\cite{Sjostrand:1987su}. There, the hard underlying event is assumed
to be a mostly incoherent sum of individual scattering processes. 
Right now, {\tt AMISIC++} is restricted to hard QCD processes and therefore 
employs the library of {\tt EXTRA\_XS}, (see Sec.~\ref{ME}). To account for a fast 
performance, however, {\tt AMISIC++} does neither evaluate matrix elements 
on-line nor uses a veto algorithm as proposed in \cite{Sjostrand:1987su}. 
Instead it pre-calculates and tabulates the appropriate differential cross 
sections and stores them to disk in the initialisation phase. This data may 
then also serve for future runs. 

\noindent
It should be noted here that {\tt AMISIC++} is in the process
of full implementation and of careful tests only. Furthermore, the description of
the soft underlying event is still lacking in {\tt Multiple\_Interactions}. 
\section{The interface to fragmentation\label{PS}}
\noindent
Having produced a number of partons in hard subprocesses -- either the 
signal process, hard particle decays, or multiple hard partonic 
interactions -- these coloured objects have to be transformed into colourless 
hadrons. The gap between the varying scales of these hard interactions and 
some universal scale connected to hadronisation is bridged by parton showers. 
Invoking the parton shower fills in further parton radiation and guarantees 
the universality of the scale, where the phenomenological hadronisation model 
sets in, and of its parameters.

\noindent
Within the {\tt SHERPA} framework, such additional emission in general 
happens during an event phase called {\tt Jet\_Evolution}. This event phase
adds blobs describing radiation of secondary partons to the list of blobs
constituting the event. To this end, all parton configurations in blobs
for signal processes, hard decays, or for multiple parton interactions
have to be analysed and modified by parton showers. The {\tt Jet\_Evolution},
thus, owns pointers to all corresponding {\tt Matrix\_\-Element\_\-Handler}s for the
definition of colour configurations and other starting conditions of the 
parton shower and to a {\tt Shower\_Handler}.
This object provides public methods that allow to initialise and perform 
showers and to insert the resulting shower blobs into the event record.
In principle, one can think of using different shower realisations, for 
instance a dipole cascade as in {\tt Ariadne} \cite{Ariadne}, an angular 
ordered shower as in {\tt Herwig} \cite{Herwig,Gieseke:2003rz}, 
or a virtuality ordered shower as in {\tt Pythia} \cite{Pythia}.
So far, in {\tt SHERPA} a virtuality-ordered shower has been 
implemented through a separate module called {\tt APACIC++} \cite{Kuhn:2000dk}. 
This module also includes the functionality needed for the merging of parton 
showers and matrix elements in the fashion of \cite{CKKW}, i.e.\ a veto on 
jets at the parton level. The implementation of other approaches that model multiple 
emission of secondary partons will not substantially change the interface 
{\tt Shower\_Handler}.\\

\noindent
From the brief description above, it is clear that the matrix elements
and the parton showers might act on different objects. In the case
realized so far, i.e.\ in the case of {\tt APACIC++}, the parton shower is 
formulated in terms of trees and knots; for a shower described in the fashion
of {\tt Ariadne} one could imagine that dipole objects are the basic terms.
Hence, in the case of {\tt APACIC++} being the parton shower generator
the {\tt Jet\_Evolution} would have to administer the translation of partons 
to knots, i.e.\ the definition of a primordial tree structure representing a 
parton configuration. This is done through suitable interfaces. The specific 
instantiation of the abstract base class {\tt Perturbative\_Interface} 
depends on the form of the matrix elements and their functionality inside 
the {\tt Matrix\_Element\_Handler}, and on the {\tt Shower\_Handler} itself. The 
application of these interfaces is mandatory for the {\tt Jet\_Evolution} 
and results in some ``merging blobs'' around the blob of the hard 
subprocess under consideration. These merging blobs are needed for the sake 
of four-momentum conservation, since secondary emission a posteriori gives a 
virtual mass to the primary on-shell partons, which has to be balanced by 
shifting the four-momenta of the primary parton ensemble. All of these 
interfaces are part of the {\tt SHERPA} framework itself rather than of the 
individual modules (such as {\tt AMEGIC++} etc.). Due to the merging algorithm, 
this interface needs to supply the possibility to calculate Sudakov weights, 
and to accept or reject parton configurations according to them. It is clear 
that a rejection necessitates a new parton configuration and, therefore, 
results in a new event to be supplied by the {\tt Matrix\_Element\_Handler}.
Correspondingly, a new {\tt Blob} is filled by the 
{\tt Signal\_Processes} event phase. However, since at the moment only two 
specific matrix element generators are available, cf.\ Sec.~\ref{ME}, only two 
realisations of the {\tt Perturbative\_Interface} exist, namely 
{\tt SimpleXS\_Apacic\_Interface} and {\tt Amegic\_Apacic\_Interface}.
 
\noindent
The former is very simple, since the library of $2\to 2$ subprocesses
is used such that additional jets are the result of the simulation of the 
radiation activity through the parton showers. Therefore, in this case, 
no veto on extra jets has to be performed inside a shower and consequently
no Sudakov form factor has to be applied. Furthermore, the colour structure 
of the partons as well as the hard scale of the subprocess can be obtained 
directly from the {\tt XS\_Base}s inside {\tt EXTRA\_XS} through simple
access methods made available to the {\tt SimpleXS\_Apacic\_Interface}.
The starting conditions for the shower are obtained in quite a 
straightforward fashion. The initial virtualities for the shower evolution 
are given by the scale of the hard subprocess, which is connected to the
maximal momentum transfer along coloured lines. The maximal opening angle of
the next emission for each parton is obtained from the angles w.r.t.\ 
to the colour connected partons in the hard $2\to 2$ process. The parton
shower is then simply initialised by filling this information into the 
trees of {\tt APACIC++}. 

\noindent
When using {\tt AMEGIC++} or any other matrix element generator involving
$2\to n$ processes with $n>3$ the situation is more complicated. In such 
cases, the $2 \to n$ configuration is reduced to a ``core'' $2 \to 2$ 
process through the $k_\perp$-cluster algorithm. To keep track of allowed
and disallowed clusterings, i.e.\ of actual Feynman rules, the clustering
follows the Feynman diagrams of the corresponding matrix element. They
are obtained through the {\tt Matrix\_Element\_Handler}. For each clustering, 
a Sudakov form factor is evaluated and attached as an extra weight (for 
details see \cite{CKKW}), which finally results in an overall weight 
for this specific parton-level event. In case it is accepted, the initial
colour structure is fixed by the colour structure of the core $2\to 2$
process, since the parton shower inherently is formulated in the large 
$N_c$-approximation. In the clustering procedure the tree structure for the
parton shower already has been constructed. It is supplemented with missing 
information (i.e.\ the starting virtualities for each parton, opening 
angles etc.) through the principle that the parton shower evolution of each 
parton is defined through the node in which it was produced first. 

\noindent
This condenses in the following algorithm: going from inner knots to the 
outer ones, in each node it is decided by the {\tt Perturbative\_Interface} 
which emerging parton is the harder, i.e.\ more energetic, one. The winner 
inherits the starting scale and angle of the decaying mother, the losers 
starting conditions are defined through the actual node. The starting conditions 
of the first four partons stem from the core $2\to 2$ subprocess.\\

\noindent
As already stated, the interface to the showers and the actual physics 
implementation are separated. Whereas the interface is located in the
{\tt Shower\_Handler}, the first physics implementation of a parton shower
is encapsulated in the independent module {\tt APACIC++}. It provides
a virtuality ordered parton shower, supplemented with angular ordering
enforced ``by hand'', similar to the one realized in {\tt Pythia}. One of
the major differences, however, is that in {\tt SHERPA}
matrix elements for arbitrary parton configurations are merged consistently
with the parton shower. This merging procedure results in constraints on the 
parton shower, which must not produce any parton emission that would have to
be interpreted as the production of an extra jet, since jet production is left
to the corresponding matrix elements. 

\noindent
The parton shower in {\tt APACIC++} is organised recursively in terms of 
binary tree structures, where the emission of an additional parton is 
understood as a branching process giving rise to another node, a {\tt Knot}, 
inside the {\tt Tree}\footnote{These trees are the only objects of 
{\tt APACIC++}, which are handed over to the {\tt Shower\_Handler} in order 
to be filled with partons subject to further emission. This process is 
triggered by the {\tt Shower\_Handler} and managed by the 
{\tt Hard\_Interface}, the class managing the access to {\tt APACIC++}}. 
In the evolution of the tree the binary branches are defined through
splitting functions, which are represented by objects of similar name,
i.e.\ by derivatives of the base class {\tt Splitting\_Function}.
These objects contain methods to determine outgoing flavours of a branching
process and their kinematics. Since in {\tt APACIC++} the parton shower
proceeds through a hit-or-miss method, functions overestimating the 
integral of a splitting function in certain boundaries and corresponding
correction weights are also included. For the incorporation of new branching 
modes, such as for the simulation of parton showers off super-symmetric 
particles, just a suitable derivative of the base class has to be added.
The sequence of branches within the parton shower is defined through Sudakov
form factors. Consequently, such objects are also used by {\tt APACIC++}.
For the description of parton showers in the initial state, backward evolution 
relying on the parton distribution functions usually is employed.
Therefore, the corresponding PDFs are handed over to {\tt APACIC++} and
used in the space-like showers and Sudakov form factors. Here, it should be 
briefly mentioned that the Sudakov form factors, in principle, provide only
the trees of branching processes. There, each node is supplemented by the 
scale, where the branching takes place, and the distribution of energies.
The corresponding determination of the actual kinematics is separated from 
the implementation of the Sudakov form factors; it is located in extra 
classes. However, once the parton shower has terminated, the tree structure
is translated back into partons. The interface, i.e.\ the {\tt Shower\_Handler},
will provide blobs with one incoming parton stemming from the hard matrix 
element, which is identified as the jet's seed, and a number of outgoing partons
exhibiting the partonic structure of the jet before hadronisation sets in.
\section{Hadronisation \& soft physics aspects\label{Soft}}
\noindent
After the parton shower described above has terminated, one is left with a
configuration of coloured partons at some low scale of the order of a few
GeV in transverse momentum. These partons, in order to match experiments,
have to be translated into white hadrons. Within {\tt SHERPA}, this
transition occurs in an event phase called {\tt Hadronisation}. This
{\tt Event\_\-Phase\_\-Handler} contains interfaces to two physics tasks
related to this phase. \\

\noindent
First of all, extracting a coloured parton from a white initial hadron 
(such as in collisions involving protons), necessitates to describe the 
colour structure of its remnant. This is achieved by the 
{\tt Beam\_\-Remnant\_\-Handler}. 

\noindent
It is clear that the coloured constituents will be colour connected to other 
partons in the final state, thus influencing properties of the event at 
hadron level. The distribution of colour over the hadron remnants is a 
tricky task, well beyond perturbation theory. This immediately implies
that phenomenological models have to be employed. For instance, one could 
assume that such a model is guided by the attempt to minimise the string 
length of the colour string spanned by the outgoing partons. Therefore, within 
{\tt SHERPA} the beam remnants arising from hadrons are currently handled 
in a naive approach. Given a list of {\tt Blob}s, all initiators 
of initial state showers are extracted and attached to a beam blob, which 
represents the breakup of the incoming hadron. Beam-\-remnant partons are added 
such that the flavour quantum numbers of the hadron are recovered step by step.
Colours are distributed in a randomised fashion, where, of course, gluons or
quarks carry two or one colour index different from zero, respectively. Again, 
these indices are distributed such that they add up to a white hadron. The 
energies of the additional parton remnants are distributed either according to 
PDFs or to a phenomenological function like the one in \cite{Sjostrand:1987su}.
Finally, all particles obtain a mild $k_\perp$-kick according to a
Gaussian distribution.\\

\noindent
The resulting final parton configuration then originates from the
perturbative event phases, i.e.\ from {\tt Signal\_\-Processes},
{\tt Hard\_\-Decays}, {\tt Multiple\_\-Inter\-actions} or {\tt Jet\_\-Evolution},
or from the beam remnants as described above\footnote{Altogether these
partons must form a colour singlet, although, if baryon-\-number violating
sub-processes are implemented, it might be difficult to recover them as 
singlets in the large $N_c$-representation inherent to event generation.}.
The {\tt Hadronisation} phase has to translate these coloured partons
into white hadrons. For this purpose, it employs its
{\tt Fragmentation\_\-Handler}, which provides an interface to phenomenological
hadronisation models.

\noindent
The {\tt Fragmentation\_Handler} first of all sorts the partons into
disconnected chains starting with a colour-triplet, such as a quark, and
ending with a parton in a colour-\-anti-\-triplet state, such as an anti-quark. 
Then, within these chains, partons are shifted to their constituent 
mass-shells, if necessary. Only then, the selected individual hadronisation 
model is invoked. This mass-shift inside the {\tt Fragmentation\_\-Handler}
guarantees the independence of the perturbative phase, which presumably is 
formulated in terms of current masses, and the non-perturbative phase with
its constituent masses. Especially for cluster-fragmentation models 
\cite{Webber:1983if} relying on the breakup of massive gluons into constituent 
quarks this is clearly advantageous. However, at the moment only the Lund string 
model \cite{Andersson:tv} is implemented as a specific hadronisation model to be 
used by the {\tt Fragmentation\_Handler}. Its implementation within {\tt Pythia} 
is accessible through a special {\tt Lund\_Fortran\_Interface} class, which 
also reads in some of the parameters needed in this model from a corresponding 
data file. In the near future, also a new version of the cluster-hadronisation
model \cite{Winter:2003tt} will be made available. 

\noindent
This model will be added as a new module, {\tt AHADIC++}, to the overall 
framework. This module just finished construction and currently is being 
tested. It performs the transition from partons to primary hadrons 
in two steps: first of all, the gluons experience a forced decay into 
colour-triplet pairs, which allows to decompose the parton sinlget chain into 
clusters. The clusters are built from one triplet--\-anti-\-triplet pair
and thus have the quantum numbers of hadrons, including those of baryons. 
In this step of cluster formation effects of soft colour reconnection are 
modelled, which is an extension to the previous versions of the cluster model 
\cite{Webber:1983if}. In the next step, the clusters decay either into lighter 
ones, or into the primary hadrons. The respective decay mode depends on the cluster 
mass and on the masses emerging for the resulting four-vectors. The distribution 
of the decay products' momenta is governed by some universal anisotropic kinematics, 
the selection of the decay mode thus reflects a constituent-\-flavour-\-dependent 
separation into a cluster and a hadron regime. There, also soft colour reconnection 
effects are taken into account. In the rare case that a primary cluster already 
is inside the hadron regime a one-particle transition is enforced. For more 
details on this model, cf. \cite{Winter:2003tt}.

\noindent
In any case, invoking the {\tt Fragmentation\_\-Handler} results in a number 
of colour singlet parton chains, each of which enters a new {\tt Blob}, 
producing a number of primordial hadrons. These hadrons may or may not decay 
further; at the moment, the subsequent hadron decays are also handled through the
{\tt Lund\_\-Fortran\_\-Interface}. In the future, however, it is envisioned
to have an extra event phase {\tt Hadron\_\-Decays} and specific interfaces. 
Each of the hadron decays is then represented
by another {\tt Blob}, allowing to reconstruct displaced vertices etc..
\section{Summary \& outlook\label{Finish}}
In this publication a proof-of-concept version of the new event-generation
framework {\tt SHERPA}, {\tt S}imluation for {\tt H}igh-{\tt E}nergy
{\tt R}eactions of {\tt PA}rticles, has been presented in its version 
1.$\alpha$. Its construction is a still on-going process, which is based 
on three programming paradigms, namely modularity, the separation of 
interface and physics implementation and a bottom-to-top approach for the 
addition of further modules. In its overall structure, {\tt SHERPA} 
reflects a typical, event-\-generator-\-inherent simulation of full events
through disjoint event phases. This lends itself to modularisation
and, therefore, {\tt SHERPA} is entirely written in the object-oriented
programming language {\tt C++}.

\noindent
So far a number of physics modules have been attached to {\tt SHERPA},
which allow users to fully simulate electron--\-positron or unresolved
photon--\-photon collisions at high energies. Also, fully hadronic 
collisions, such as, e.g., proton--\-anti-proton or proton--\-proton reactions,
can be simulated. In the description of such events, however, some
features, for instance the soft underlying event, are still lacking
or basically not tested yet. In all cases considered so far, {\tt SHERPA}
proved to be flexible and to live up for all demands. More tests
and the inclusion of further, nearly ready physics modules, such as 
a new version of the cluster hadronisation, hard decays of unstable 
heavy particles, or an underlying event model, will be in the focus of
future work. 

\noindent
{\tt SHERPA} can be obtained through the downloads section of:
\begin{verbatim}
http://www.physik.tu-dresden.de/~krauss/hep/index.html
\end{verbatim}

\section*{Acknowledgements}
F.K.\ wants to acknowledge financial support by the EC 5th Framework 
Programme under contract number HPMF-CT-2002-01663. Further financial
support by BMBF, DFG and GSI is gratefully acknowledged. \\
The authors are grateful for fruitful discussions with 
Stefan Gieseke, Klaus Ha\-macher, Hendrik Hoeth, Leif L{\"o}nn\-blad, Al\-berto Ri\-bon, 
Ger\-hard Soff, Philip Stevens, and Bryan Webber.
Also, the authors owe a great deal to users of {\tt SHERPA}, which have 
reported on bugs and shortcomings, in particular Claude Charlot, Alessio Ghezzi, 
Hendrik Hoeth, Huber Nieto, and Thorsten Weng\-ler. 
Without all this help such a task would be unsurmountable.

\end{document}